\documentclass[aps,twocolumn,prb,floatfix,showpacs,lengthcheck,citeautoscript,superscriptaddress,longbibliography]{revtex4-2}
\usepackage{amsmath}
\usepackage{amssymb}
\usepackage{graphicx}
\usepackage{bm}
\usepackage{float}
\usepackage{mathtools}
\usepackage{color,soul}
\usepackage{times}
\usepackage{upgreek}
\usepackage{MnSymbol}
\usepackage{verbatim}
\usepackage[bottom]{footmisc}
\usepackage{bbold}

\usepackage[dvipsnames,svgnames,table]{xcolor}
\usepackage{hyperref}
\usepackage{fancyhdr}
\usepackage[english]{babel}
\usepackage[caption=false]{subfig}

\hypersetup{
pdfstartview={FitH},
colorlinks=true,    
linkcolor=NavyBlue, 
citecolor=Blue,   
filecolor=NavyBlue, 
urlcolor=NavyBlue   
}

\newcommand{\bea}{\begin{eqnarray}}
\newcommand{\eea}{\end{eqnarray}}
\newcommand{\be}{\begin{equation}}
\newcommand{\ee}{\end{equation}}

\begin{document}
\title{Imaging chiral Andreev reflection in the presence of Rashba spin-orbit coupling}
\author{Lucila {Peralta Gavensky}}
\affiliation{Centro At{\'{o}}mico Bariloche and Instituto Balseiro,
Comisi\'on Nacional de Energ\'{\i}a At\'omica (CNEA)- Universidad Nacional de Cuyo (UNCUYO), 8400 Bariloche, Argentina}
\affiliation{Instituto de Nanociencia y Nanotecnolog\'{i}a (INN-Bariloche), Consejo Nacional de Investigaciones Cient\'{\i}ficas y T\'ecnicas (CONICET), Argentina}

\author{Gonzalo Usaj}
\affiliation{Centro At{\'{o}}mico Bariloche and Instituto Balseiro,
Comisi\'on Nacional de Energ\'{\i}a At\'omica (CNEA)- Universidad Nacional de Cuyo (UNCUYO), 8400 Bariloche, Argentina}
\affiliation{Instituto de Nanociencia y Nanotecnolog\'{i}a (INN-Bariloche), Consejo Nacional de Investigaciones Cient\'{\i}ficas y T\'ecnicas (CONICET), Argentina}

\author{C. A. Balseiro}
\affiliation{Centro At{\'{o}}mico Bariloche and Instituto Balseiro,
Comisi\'on Nacional de Energ\'{\i}a At\'omica (CNEA)- Universidad Nacional de Cuyo (UNCUYO), 8400 Bariloche, Argentina}
\affiliation{Instituto de Nanociencia y Nanotecnolog\'{i}a (INN-Bariloche), Consejo Nacional de Investigaciones Cient\'{\i}ficas y T\'ecnicas (CONICET), Argentina}

\begin{abstract}
In this work, we theoretically study transverse magnetic focusing in a two-dimensional electron gas with strong Rashba spin-orbit interaction when proximitized along its edge with a superconducting contact. The presence of superconducting correlations leads to the emergence of chiral Andreev edge states which---within this weak magnetic field regime---may be pictured as states following semiclassical skipping orbits with alternating electron-hole nature. The spin-orbit induced splitting of the Fermi surface causes these carriers to move along cyclotron orbits with different radii, allowing for their spatial spin separation. When Andreev reflection takes place at the superconducting lead, scattered carriers flip both their charge and spin, generating distinguishable features in the transport properties of the device. In particular, we report a notable enhancement of the separation between the spin-split focal points, which scales linearly with the number of Andreev scattering events at the anomalous terminal. We support our results by calculating conductance maps to arbitrary points in the sample that provide a complete image of the ballistic electron-hole cyclotron paths.
\end{abstract} 
\maketitle
\section{Introduction}

Chiral Andreev edge states are one-way hybrid electron-hole modes that propagate along the interface of a Hall sample and a superconductor (SC) but remain bounded in the perpendicular direction due to both the magnetic field and the superconducting gap  confinement. The hybrid nature of this distinctive type of edge state is rooted in the proximity effect: the conventional edge states living at the boundary of the Hall region acquire superconducting correlations via successive Andreev reflections at the interface with the anomalous contact~\citep{Klapwijk2004,Hoppe2000,Giazotto2005}, ultimately leading to a coherent superposition of propagating modes with opposite charge. Since backscattering is not allowed, the carriers circulate chirally in the direction determined by the external magnetic field. 

Transparent interfaces between superconducting alloys and samples in the Hall regime are now within experimental reach~\citep{Wan2015,Lee2017,Park2017,Zhi2019}, providing the condensed matter community with an exciting playground for probing transport phenomena occurring along the boundary of these two phases of matter. Indeed, interference of chiral Andreev edge states in the quantum Hall (QH) regime has just been reported~\citep{Zhao2020}. Several theoretical studies have focused on the large magnetic field limit---more specifically, on the mechanisms by means of which an edge mediated current may flow in a SC-QH-SC Josephson junction with only one filled Landau level~\citep{Ma1993,Stone2011,vanOstaay2011,Alavirad2018,PeraltaGavensky2020,PeraltaGavensky2021}. Substantial experimental advances were also made in this direction~\citep{Amet2016,Guiducci2018,Draelos2018,Seredinski2019}. 

On the other hand, weak magnetic field regimes with large filling fractions have been scarcely analyzed~\citep{Haugen2011,Polinak2006,Kormanyos2007}. Within this range of fields, the semiclassical skipping orbits of electronic and hole-like states were only recently imaged in a magnetic focusing setup in Ref.~\cite{Bhandari2020}. By using scanning gate microscopy techniques \cite{Aidala2007}, the Andreev reflected carriers were successfully detected while following ballistic cyclotron paths in a graphene sample.

\begin{figure}[b]
\includegraphics[width=0.95\columnwidth]{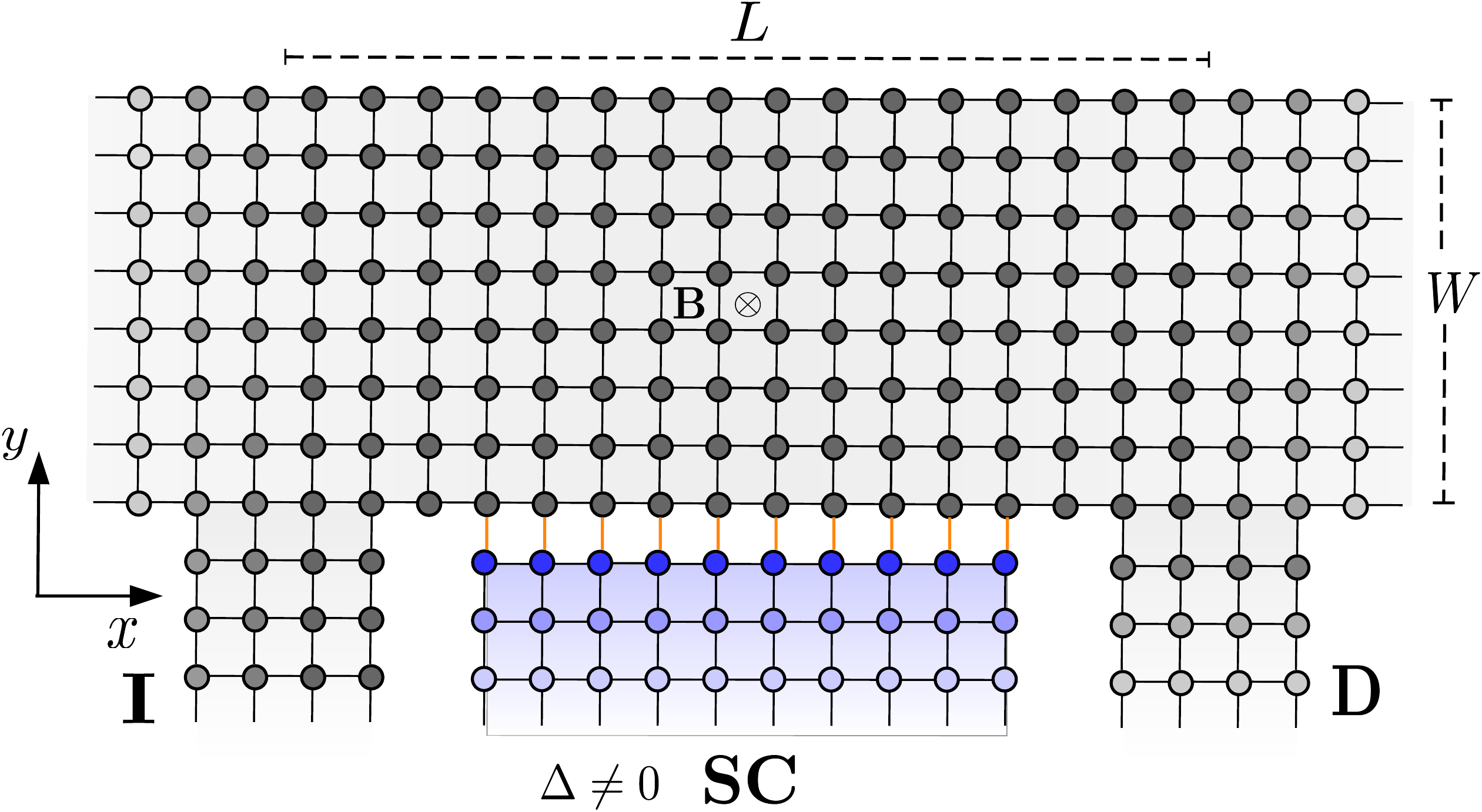}
\caption{Illustration of the geometry used to study the transverse magnetic focusing. Two normal contacts, the source lead I and the drain lead D are attached to a ribbon of width $W$ which is subjected to an external magnetic field $\bm{B}=-B\bm{\hat{z}}$. These terminals are at a fixed distance $L$ from each other. A third superconducting terminal (SC) with a non-vanishing order parameter $\Delta$  is placed in between the two normal contacts.}
\label{fig1}
\end{figure}
In this work, we pose the question of what would happen if such magnetic focusing experiments were to be performed in two-dimensional electron gases (2DEGs) with significant spin orbit (SO) coupling, as the ones used in Refs.~\citep{Wan2015,Zhi2019}.  These experiments consist on the injection of electrons into a 2DEG through a voltage biased contact. In the presence of a small magnetic field, the carriers follow a skipping orbit trajectory which is essentially determined by the shape of the Fermi surface, eventually focusing at certain points along the edge ~\cite{Houten1989,Beenakker1991a}. By tuning the magnetic field, the focal distance can be adjusted to match the one of a detector lead. In this way, the collected carriers give rise to an abrupt change in the conductance between the two aforementioned contacts, or the appearance of a voltage~\cite{Potok2002,Potok2003}. Transverse magnetic focusing in SO coupled systems has been useful technique to study spatial spin separation in mesoscopic devices~\cite{Rokhinson2004,Dedigama2006,Heremans2007,Li2012,Lo2017}. It is well known that the presence of two spin-split Fermi surfaces leads to the existence of two different cyclotron paths which translates into a splitting of the focusing spectrum \cite{Usaj2004,Reynoso2007,Zuelicke2007,Reynoso2008,Kormanyos2010}. We here report how this well-established phenomenon is modified when allowing carriers to Andreev-reflect at an intermediate extended superconducting terminal as depicted in Fig.~\ref{fig1}. To this end, we perform numerical simulations of the conductance between the source lead I and the drain lead D as a function of the external magnetic field. The focusing peaks show clear signatures of Andreev reflection at the superconducting terminal. On the one hand, the sign of the conductance reveals if the arriving particle is an electron or a hole, a fact which depends on the number of bounces at the anomalous lead. More interestingly, we notice that the Andreev scattered particles preserve their cyclotron radius due to the mixing of electronic and hole states with opposite spin, making the spin-splitting of the focusing peaks to be enhanced with respect to a conventional focusing experiment. 

The work is organized as follows: in Section \ref{II} we present the Hamiltonian model for the 2DEG, the normal and superconducting leads along with their discretized tight-binding version. In Section \ref{III} we present the numerical results of the conductance between the two normal leads as a function of the external magnetic field, the strength of the spin-orbit interaction and the transparency between the 2DEG and the superconducting terminal. To better illustrate our findings we also show color maps of the electron and hole transmissions to arbitrary points in the sample. This allows for a complete imaging of the chiral Andreev semiclassical edge states. Finally, in Section \ref{IV} we present a summary of our main results and some concluding remarks.

\section{Model Hamiltonian}\label{II}
The Hamiltonian of a 2DEG with Rashba spin-orbit interaction is given by
\begin{equation}
\hat{H} = \frac{1}{2m^{*}}(\hat{\Pi}_x^2 + \hat{\Pi}_y^2)+\frac{\alpha}{\hbar}(\hat{\Pi}_x\sigma_y - \hat{\Pi}_y\sigma_x)-\frac{1}{2}g\mu_B\sigma_z B\,,
\label{HDEG}
\end{equation}
where $\hat{\Pi}_{\eta}=\hat{p}_{\eta}+e A_{\eta}/c$ with $\hat{p}_{\eta}$ the momentum and $A_{\eta}$ the vector potential along the $\eta$ direction, $m^{*}$ stands for the effective mass and $\alpha$ for the Rashba coupling parameter. The Pauli matrices $\{\sigma_{\eta}\}$ act in spin-space. The magnetic field threading the sample is given by $\bm{B}=-B\bm{\hat{z}}$ and $g$ is the effective gyromagnetic factor. The numerical simulations are carried out performing a space discretization of the model, so that the tight-binding version can be written as $\hat{H}=\hat{H}_0 + \hat{H}_R$ with
\begin{eqnarray}
\hat{H}_0 &=& \sum_{\bm{r}\sigma}\varepsilon_{\sigma}c^{\dagger}_{\bm{r},\sigma}c^{}_{\bm{r},\sigma}\\
\notag
&-&t\sum_{\bm{r}\sigma}\left(e^{-i\frac{2\pi\Phi}{\Phi_0}\frac{y}{a_0}}c^{\dagger}_{\bm{r},\sigma}c^{}_{\bm{r}+a_0\bm{\hat{x}},\sigma}+c^{\dagger}_{\bm{r},\sigma}c^{}_{\bm{r}+a_0\bm{\hat{y}},\sigma}+\text{H.c.}\right)\,,
\end{eqnarray}
and
\begin{eqnarray}
\hat{H}_R &=& -\lambda\sum_{\bm{r}}\Big[i\Big(c^{\dagger}_{\bm{r},\uparrow}c^{}_{\bm{r}+a_0\bm{\hat{y}},\downarrow} + c^{\dagger}_{\bm{r},\downarrow}c^{}_{\bm{r}+a_0\bm{\hat{y}},\uparrow}\Big)\\
\notag
&-& e^{-i\frac{2\pi\Phi}{\Phi_0}\frac{y}{a_0}}\Big(c^{\dagger}_{\bm{r},\uparrow}c^{}_{\bm{r}+a_0\bm{\hat{x}},\downarrow}-c^{\dagger}_{\bm{r},\downarrow}c^{}_{\bm{r}+a_0\bm{\hat{x}},\uparrow}\Big)\Big]+\text{H.c.}\,.
\end{eqnarray}
Here $c^{\dagger}_{\bm{r},\sigma}$ creates an electron at the 2DEG's site $\bm{r}=x\bm{\hat{x}}+y\bm{\hat{y}}$ with spin $\sigma$, $a_0$ is the lattice spacing, $\varepsilon_{\sigma}=4t - \mu - \sigma g\mu_B B/2$, $t=\hbar^2/2m^{*}a_0^2$, $\lambda = \alpha/2 a_0$, $\Phi = -B a_0^2$ is the flux per plaquette and $\Phi_0 = hc/e$ is the normal flux quantum. The orbital effect of the magnetic field is included via the Peierls substitution. We have used a Landau gauge where the vector potential $\bm{A}= B y \bm{\hat{x}}$. The two lateral normal contacts I and D are described by narrow stripes of $N_0$ sites with, for simplicity, no spin-orbit coupling ($\alpha=0$). We choose to gate voltage these terminals so that they have a single active channel at the Fermi level.
The superconducting terminal is modeled as a square lattice with a Hamiltonian given by 
\begin{eqnarray}
\hat{H}_{SC} &=& \sum_{\bm{r}\sigma}(4t -\mu)s^{\dagger}_{\bm{r},\sigma}s^{}_{\bm{r},\sigma}-t\sum_{\bm{r}\sigma}\Big(s^{\dagger}_{\bm{r},\sigma}s^{}_{\bm{r}+a_0\bm{\hat{x}},\sigma}\\
\notag
&+&s^{\dagger}_{\bm{r},\sigma}s^{}_{\bm{r}+a_0\bm{\hat{y}},\sigma}+\text{H.c.}\Big) -\Delta\sum_{\bm{r}}\Big(s^{\dagger}_{\bm{r},\uparrow}s^{\dagger}_{\bm{r},\downarrow}+s^{}_{\bm{r},\downarrow}s^{}_{\bm{r},\uparrow}\Big)\,,
\end{eqnarray}
where $s^{\dagger}_{\bm{r},\sigma}$ creates an electron at the superconductor's site $\bm{r}$ with spin $\sigma$ and a local pairing potential $\Delta$ has been included to simulate superconductivity. The end sites of this terminal are coupled with the 2DEG with a tunneling matrix element $\gamma=t$ unless otherwise stated. From now on we choose $a_0=5\,\text{nm}$ and $m^{*}=0.055\,m_0$ with $m_0$ the electron mass. The superconducting gap is taken to be $\Delta=1\,\text{meV}$. Regarding the geometry, we selected the normal leads to have $N_0=21$ sites with a fixed distance between each other of $L=249a_0$.  The superconducting terminal has $N_s=190$ sites and is placed symmetrically in between the source and the drain leads. 

When applying a bias voltage in the injector lead I while leaving the rest of the system to ground, the conductance between I and the drain lead D can be obtained as
\begin{equation}
G=\frac{e^2}{h}\sum_{\sigma\sigma'} (T^{DI}_{e\sigma,e\sigma'}-T^{DI}_{h\sigma,e\sigma'}),
\label{eqG}
\end{equation}
where
\begin{eqnarray}
\label{Tee}
T^{DI}_{e\sigma,e\sigma'} &=& \text{Tr}\Big[\Gamma^{D}_{e\sigma} \mathcal{G}^{r} \Gamma^{I}_{e\sigma'}\mathcal{G}^{a}\Big]\\
\label{The}
T^{DI}_{h\sigma,e\sigma'} &=& \text{Tr}\Big[\Gamma^{D}_{h\sigma} \mathcal{G}^{r} \Gamma^{I}_{e\sigma'}\mathcal{G}^{a}\Big].
\end{eqnarray}
Here $T_{e\sigma,e\sigma'}^{DI}$ ($T_{h\sigma,e\sigma'}^{DI}$) is the transmission coefficient of an electron with spin $\sigma'$ from lead I as an electron (hole) of spin $\sigma$ to lead D and $\mathcal{G}^{r}$ ($\mathcal{G}^{a}$) is the retarded (advanced) Green's function of the sample. We used a Bogoliubov-de Gennes spinor basis as a way of keeping track of electronic, hole and spin sectors so that the spatial elements of the retarded propagator are written as
\begin{equation}
\mathcal{G}_{\bm{r},\bm{r}'}^{r}(\omega)\!=\! -i\!\int\!d(t-t')\theta(t-t') e^{i\omega(t-t')}\langle\{\!\hat{\Psi}(\bm{r},t),\hat{\Psi}^{\dagger}(\bm{r'},t')\!\}\rangle,    
\end{equation}
with $\hat{\Psi}^{\dagger}(\bm{r},t)=(c^{\dagger}_{\bm{r},\uparrow}, c^{\dagger}_{\bm{r},\downarrow}, c^{}_{\bm{r},\downarrow}, -c^{}_{\bm{r},\uparrow})$. The elements of the advanced Green's function may be obtained in a similar fashion. The coupling matrices of each lead $m=I,D$ are defined as $\Gamma^{m}=i(\Sigma_{m}^{r}-\Sigma_{m}^{a})$ where $\Sigma_{m}^{r}$ and $\Sigma_{m}^{a}$ are the retarded and advanced self-energies of the $m$-th terminal~\citep{Datta1995}. 
All these frequency dependent quantities are evaluated at $\omega=0$ (in our formulation, the Fermi energy scale is given by $\mu$, which we choose to be $\mu=20\,\text{meV}$).

The calculations are performed by considering the 2DEG to be an infinite ribbon of width $W=500a_0$, so that the propagators of the system without contacts can be calculated by Fourier transforming along the $x$ direction. The self-energies of the normal and superconducting leads are then simply included by using the Dyson equation of motion. 
\section{Numerical results}\label{III}
\begin{figure}[b]
\includegraphics[width=0.95\columnwidth]{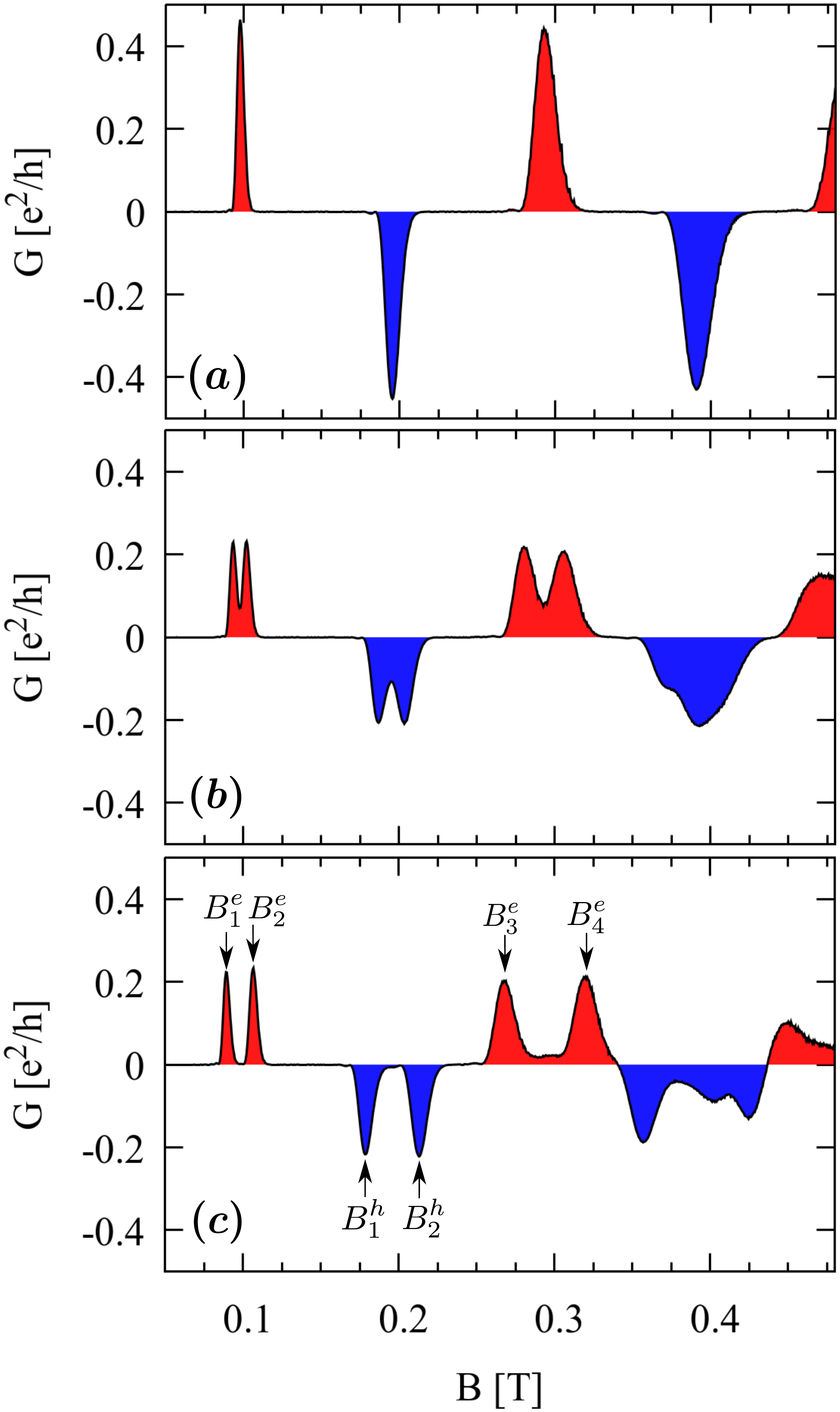}
\caption{Conductance between the source lead I and the drain lead D as obtained from Eq.~\eqref{eqG} as a function of the external magnetic field. The parameters are the ones described in Section~\ref{II}. In panel $(a)$ $\alpha=0$, $(b)$ $\alpha=10\,\text{meV nm}$ and $(c)$ $\alpha=20\,\text{meV nm}$.}
\label{fig2}
\end{figure}
\begin{figure}[t]
\includegraphics[width=0.95\columnwidth]{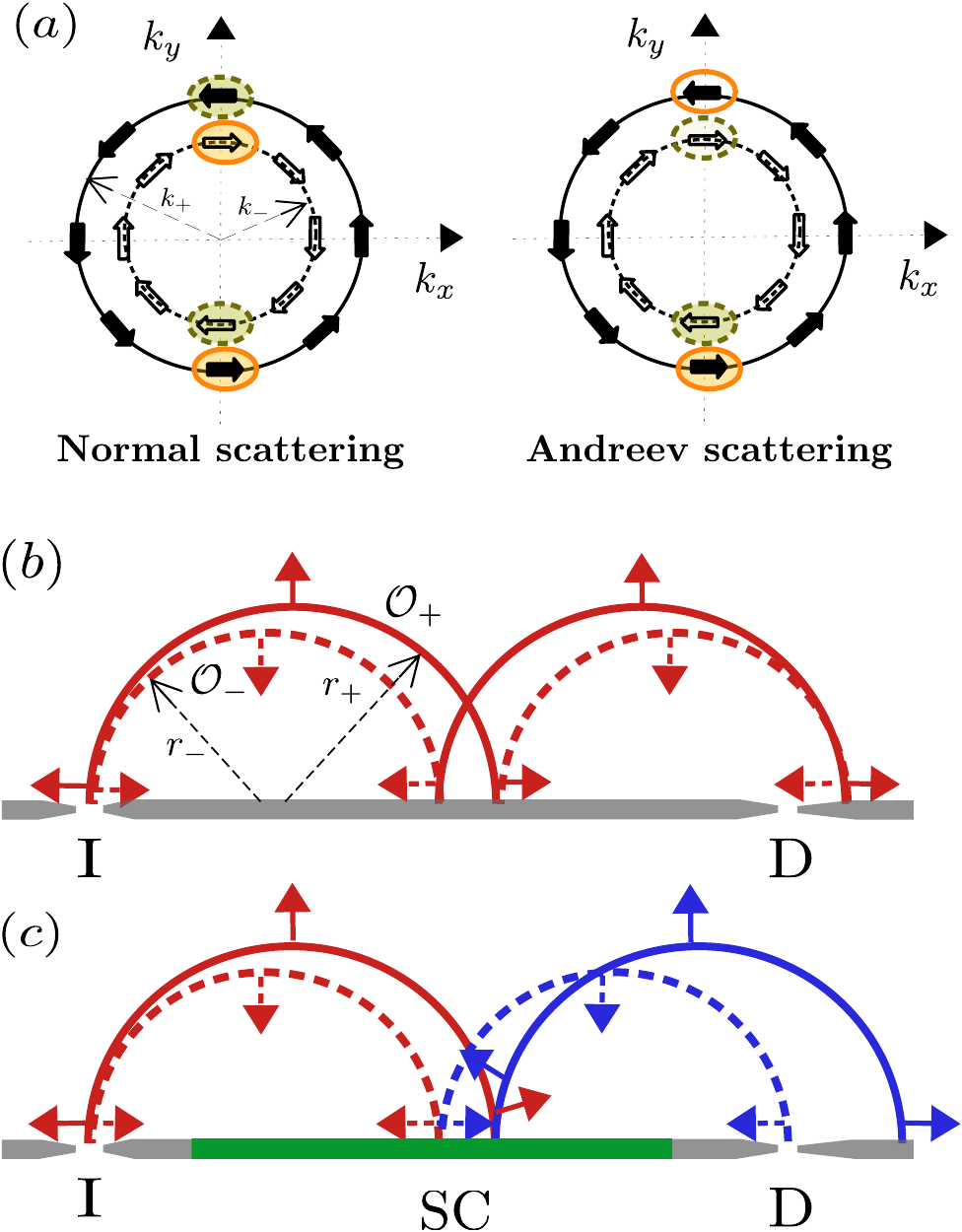}
\caption{(a) Fermi surface of a 2DEG with Rashba spin-orbit coupling and negligible Zeeman splitting. The arrows indicate the in-plane spin alignment of the eigenfunctions of Eq.~\eqref{HDEG} in the semiclassical limit. On the left hand side we show the equal-spin electron pairs (solid and dashed circles) that are involved in a normal scattering event at the edge of the sample. On the right hand side we partner the Cooper pairs that are transferred at each Andreev scattering event. The hollow circles indicate the spin of the missing carrier which is left behind as a hole-like excitation. (b) Semiclassical trajectories with normal scattering at the edge of the sample. (c) Semiclassical trajectories with Andreev reflection at an intermediate superconducting terminal (green slice). Red curves indicate electron-like carriers while the blue ones are hole-like.}
\label{fig3}
\end{figure}

We show in Fig.~\ref{fig2} the conductance $G$ as obtained from Eq.~\eqref{eqG} for different values of the spin-orbit coupling. For $\alpha=0$ [Fig.~ \ref{fig2}(a)] a series of equidistant peaks with alternating sign can be appreciated. A simple semiclassical picture explains this behavior: a well-defined peak develops whenever the condition $2r_0 N = L$ holds, with $r_0$ the cyclotron radius and $N$ an integer that labels the peak number. Recall that within this setup normal incidence along the edge is ensured, making the skipping orbits to be well described by semicircles: the drain lead matches a focal point whenever its distance from the injecting contact ($L$) becomes commensurate with the diameter of the orbit ($2r_0$). For an odd $N$ there are zero or an even number of bounces at the superconducting lead along the ballistic path, resulting in a positive (electron-like) conductance. Conversely, the even peaks are dominated by a hole-like transmission caused by an odd number of Andreev reflections at the superconductor, leading to a negative depletion of the conductance. In the experiment of Ref.~\citep{Bhandari2020} the first two peaks of Fig.~\ref{fig2}(a) were successfully detected. The cyclotron radius can be obtained as $r_0 = v_F/\omega_c$, with $v_F = \sqrt{2\mu/m^{*}}$ the Fermi velocity and $\omega_c = e B/m^{*}c$ the cyclotron frequency. With the parameters chosen in Sec.~\ref{II}, this would lead to a positive (odd $N$) or negative (even $N$) enhancement of the conductance at $B^{(N)}=N B_0$ with $B_0\simeq 0.09\,$T, in good agreement with the numerical data.
As the spin-orbit coupling $\alpha$ is increased [Figs.~\ref{fig2}(b) and (c)], \textit{all} the focusing peaks split in two. Notably, the splitting increases linearly with the peak number as $\Delta B^{(N)} = N \Delta B^{(1)}$, with $\Delta B^{(1)}$ the splitting of the first peak.  This stands in contrast with the well-known conductance spectra of focusing setups without superconducting terminals, where the odd peaks are expected to be split while the even ones are not~\citep{Usaj2004,Rokhinson2004}. To better understand these phenomena, it is useful to bear in mind the spin-texture of the Fermi surface defined by Eq.~\eqref{HDEG}. With large filling fractions and negligible Zeeman splitting, the eigenfunctions at a given energy have their spin laying in the $x$-$y$ plane as depicted in Fig.~\ref{fig3}(a). The two disconnected parts of the Fermi surface are characterized by the wavevectors
\begin{equation}
k_{\pm} = \sqrt{\frac{2\mu m^{*}}{\hbar^2} +\left(\frac{m^{*}\alpha}{\hbar^2}\right)^2}\pm \frac{m^{*}\alpha}{\hbar^2}.
\end{equation}
In real space, this leads to different classical orbit radii \cite{Usaj2004,Reynoso2007,Zuelicke2007,Reynoso2008,Kormanyos2010} coexisting within the same energy
\begin{equation}
r_{\pm} = l_B^2 k_{\pm} =\sqrt{r_0^2+\left(\frac{\alpha}{\hbar\omega_c}\right)^2 }\pm \frac{\alpha}{\hbar\omega_c},
\end{equation}
which are correspondingly associated to the two cyclotron trajectories $\mathcal{O}_{\pm}$ illustrated in Fig.~\ref{fig3}(b). Here $l_B=\sqrt{\hbar c/e B}$ stands for the magnetic length. The spin-texture suggests that, in an adiabatic picture, an electron injected at lead I with positive momentum along the $\bm{\hat{y}}$ axis and spin along the positive (negative) $\bm{\hat{x}}$ direction would undergo a precession which is well described by the $\mathcal{O}_{-}$ ($\mathcal{O}_{+}$) path, eventually reaching the edge of the sample with its spin pointing along $-\bm{\hat{x}}$ ($\bm{\hat{x}}$). The first two conductance maximums indicated in Fig.~\ref{fig2}(c) as $B_1^{e}$ and $B_2^{e}$ are obtained when the magnetic field is respectively tuned in such a way that $2 r_{-} = L$ and $2 r_{+}=L$, so that electrons are adequately collected at the D detector without intermediate scattering events. Their splitting is then determined by 
\begin{equation}
\Delta B^{(1)}=B_2^{e}-B_1^{e}= \frac{2\hbar c}{eL}(k_{+}-k_{-})  =4 \frac{ m^{*} c}{L \hbar e}\alpha.
\end{equation}
For the parameters of Fig.~\ref{fig2}(c), the above expression leads to $\Delta B^{(1)} = 17\,$mT, correctly capturing the splitting observed in the numerical simulation. If during the ballistic path normal scattering takes place at the edge of the sample, the two orbits $\mathcal{O}_{-}$ and $\mathcal{O}_{+}$ mix due to spin conservation: scattered electrons can only be reflected to their equal-spin partner, correspondingly encircled in the left side of Fig.~\ref{fig3}(a) with solid or dashed lines. Such spin-preserving scattering gives rise to a skipping trajectory along the edge where states with a large orbital radius are reflected onto states with a small orbital radius and vice versa, as shown in Fig.~\ref{fig3}(b). This effect ultimately leads to a unique unpolarized second focusing peak in conventional magnetic focusing experiments~\citep{Usaj2004}. On the other hand, when a superconducting lead is attached halfway between the lateral normal contacts, Andreev reflection takes place at the edge of the sample. As illustrated in Fig.~\ref{fig3}(c),  Andreev-scattered carriers bear not only an opposite charge but also an opposite spin, so that a singlet Cooper pair is transferred between the sample and the superconductor at each bounce. As sketched with the encircled pairs on the right hand side of Fig.~\ref{fig3}(a), the superconducting terminal takes the incident electron and another one with opposite spin and quasimomentum so that the corresponding hole-like excitation which is left behind follows the motion of the missing electron. Interestingly, an injected electron with a given spin along the $\bm{\hat{x}}$ direction will then follow a cyclotron motion with alternating electron-hole character but with a well-defined cyclotron radius on account of the conservation of the band index $\pm$. This eventually produces an enhanced separation of the subsequent peaks in the conductance spectra. In particular, the illustration of Fig.~\ref{fig3}(c) represents the condition $4 r_{-}=L$ where the conductance reaches the first minimum at $B_1^{h}$ [see Fig.~\ref{fig2}(c)]. Within this picture, it is clearly seen that the separation between the first and the second hole-like dips $\Delta B^{(2)}=B_2^{h}-B_1^{h}$ will be twice $\Delta B^{(1)}$. With each additional Andreev scattering event at the superconductor, the relative path difference between the $\mathcal{O}_{+}$ and the $\mathcal{O}_{-}$ trajectories gets linearly increased, explaining the amplified splitting observed in Fig.~\ref{fig2}(c).
\begin{figure}[b]
\includegraphics[width=0.95\columnwidth]{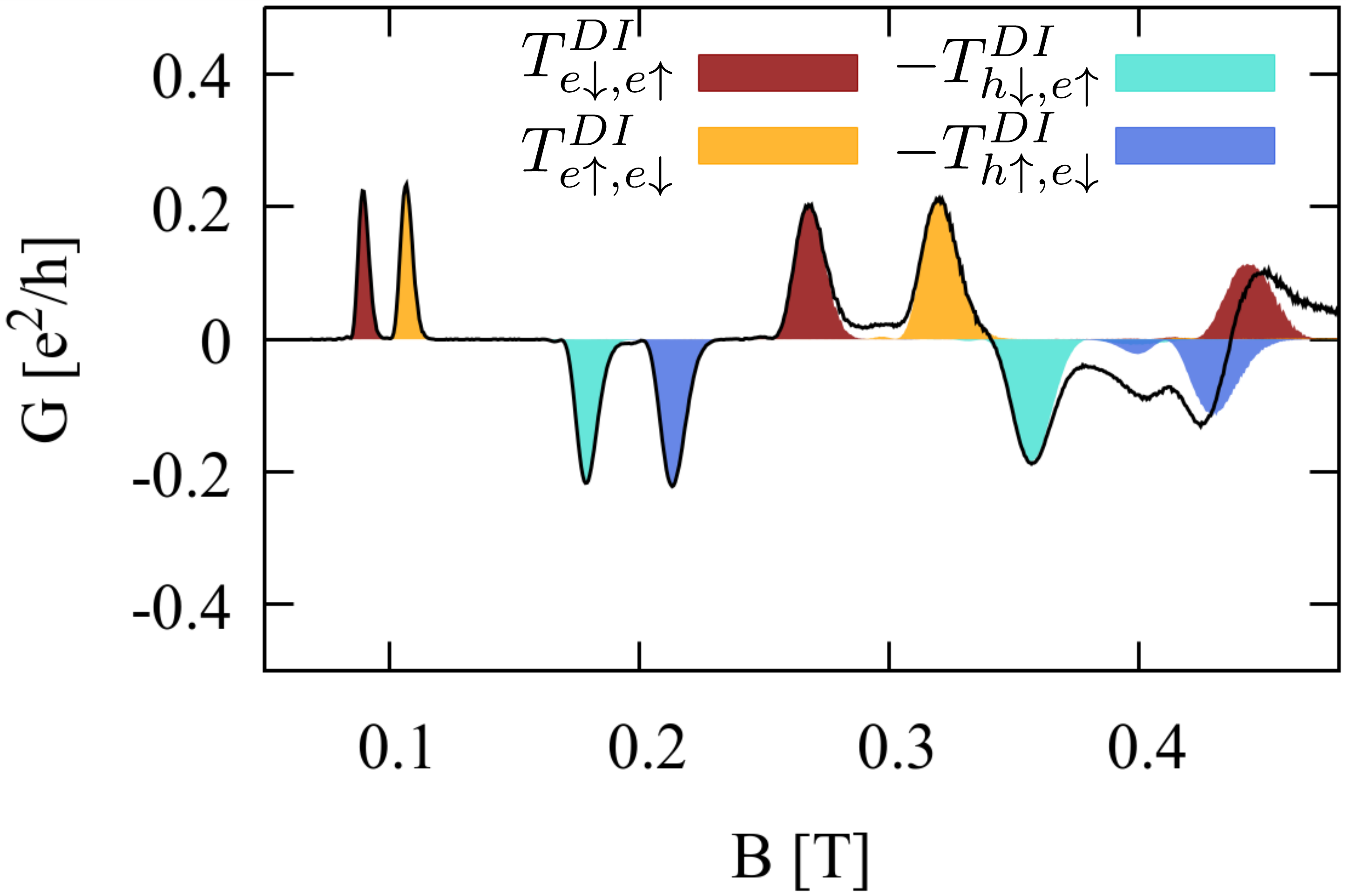}
\caption{Partial contributions of the electron and hole transmissions that dominate the conductance of Fig.~\ref{fig2}(c). Here the spin quantization axis is taken along the $\bm{\hat{x}}$ direction. The transmission coefficients are obtained from Eqs.~\eqref{Tee} and ~\eqref{The}.}
\label{fig4}
\end{figure}

\begin{figure*}[t]
\includegraphics[width=0.9\textwidth]{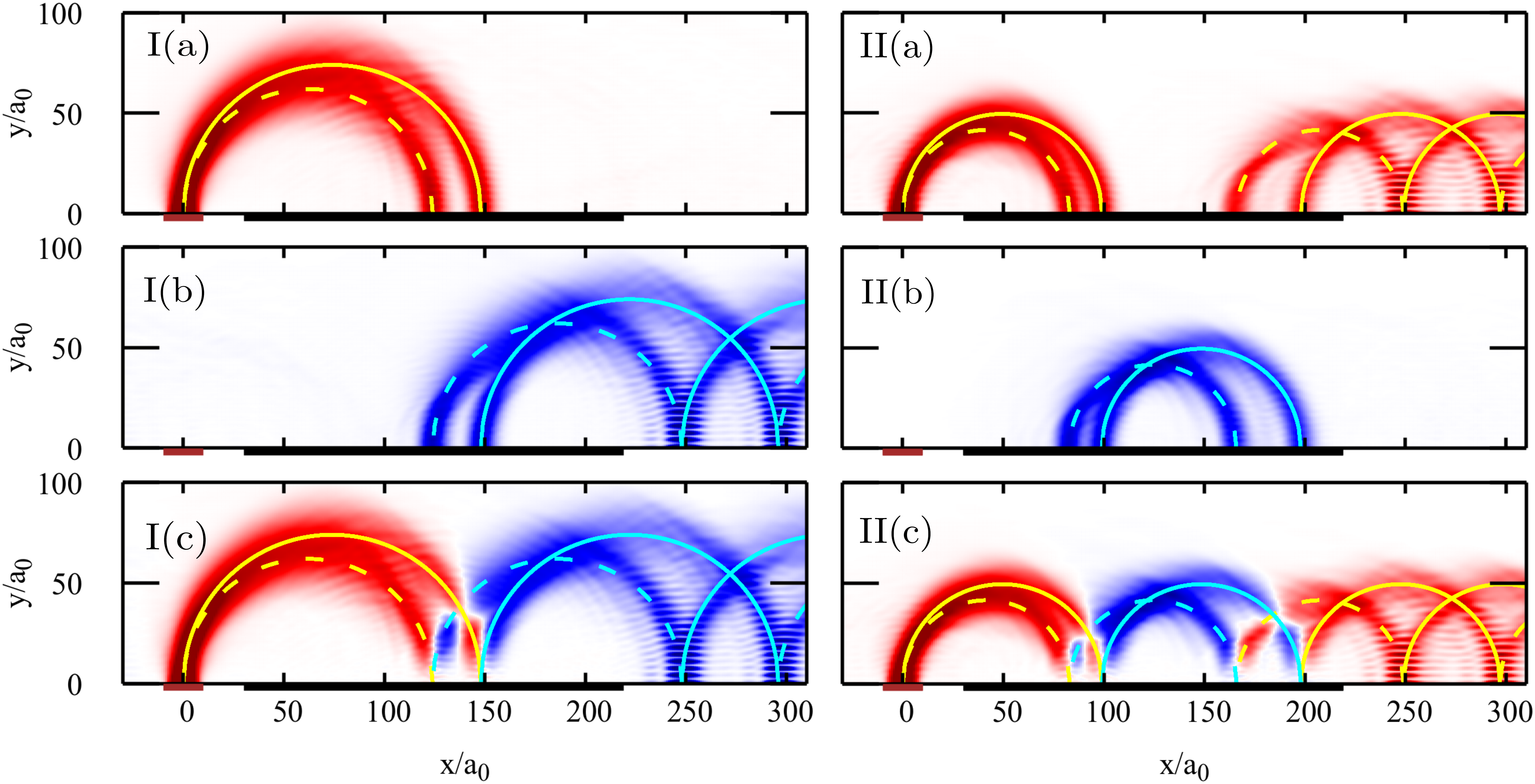}
\caption{Density plots of different transmission coefficients from the injector lead I to arbitrary points in the 2DEG. The color scale is in arbitrary units, with red being a positive conductance and blue a negative one. The spin-orbit coupling parameter is taken to be $\alpha=20\,\text{nm eV}$ as in Fig.~\ref{fig2}(c). In panels I(a)-(c) the magnetic field is chosen to be $B_1^{h}=0.179\,$T and in panels II(a)-(c) $B_3^{e}=0.268\,$T. In panels I(a) and II(a) we show the electron-electron contribution to the conductance ($\sum_{\sigma\sigma'}T^{\bm{r}I}_{e\sigma,e\sigma'}$) and, in I(b) and II(b), the electron-hole transmission with its corresponding sign ($-\sum_{\sigma\sigma'}T^{\bm{r}I}_{h\sigma,e\sigma'}$). In panels I(c) and II(c) the total conductance is shown. The position of the source lead I is delimited with a brown line and the region where the superconducting terminal is attached is indicated with a black one. Dashed and solid lines indicate the expected semiclassical orbits with a small or a large radius, respectively.}
\label{fig5}
\end{figure*}
In Fig.~\ref{fig4}, we show the partial contributions that dominate the conductance of Fig.~\ref{fig2}(c), namely the electron-electron and electron-hole transmission coefficients [see Eqs.~\eqref{Tee} and ~\eqref{The}] that involve a spin flip along the $\bm{\hat{x}}$ direction.  These results are consistent with the semiclassical picture discussed above: the first and second electronic peaks are almost entirely polarized along the negative and positive $x$-axis respectively, a clear manifestation of the spin rotation effect induced by the coupling of spin and momenta along the cyclotron motion. The second pair of hole-like dips can be understood as the outcome of one electronic and one hole-like spin rotation connected by an intermediate Andreev reflection [see Fig.~\ref{fig3}(c)]. The collected positively charged carriers will then arrive with a spin which is reversed relative to the one of the negatively charged injected carriers. Due to the linear amplification of the splitting, the peaks begin to overlap at larger magnetic fields.

\begin{figure}[t]
\includegraphics[width=0.95\columnwidth]{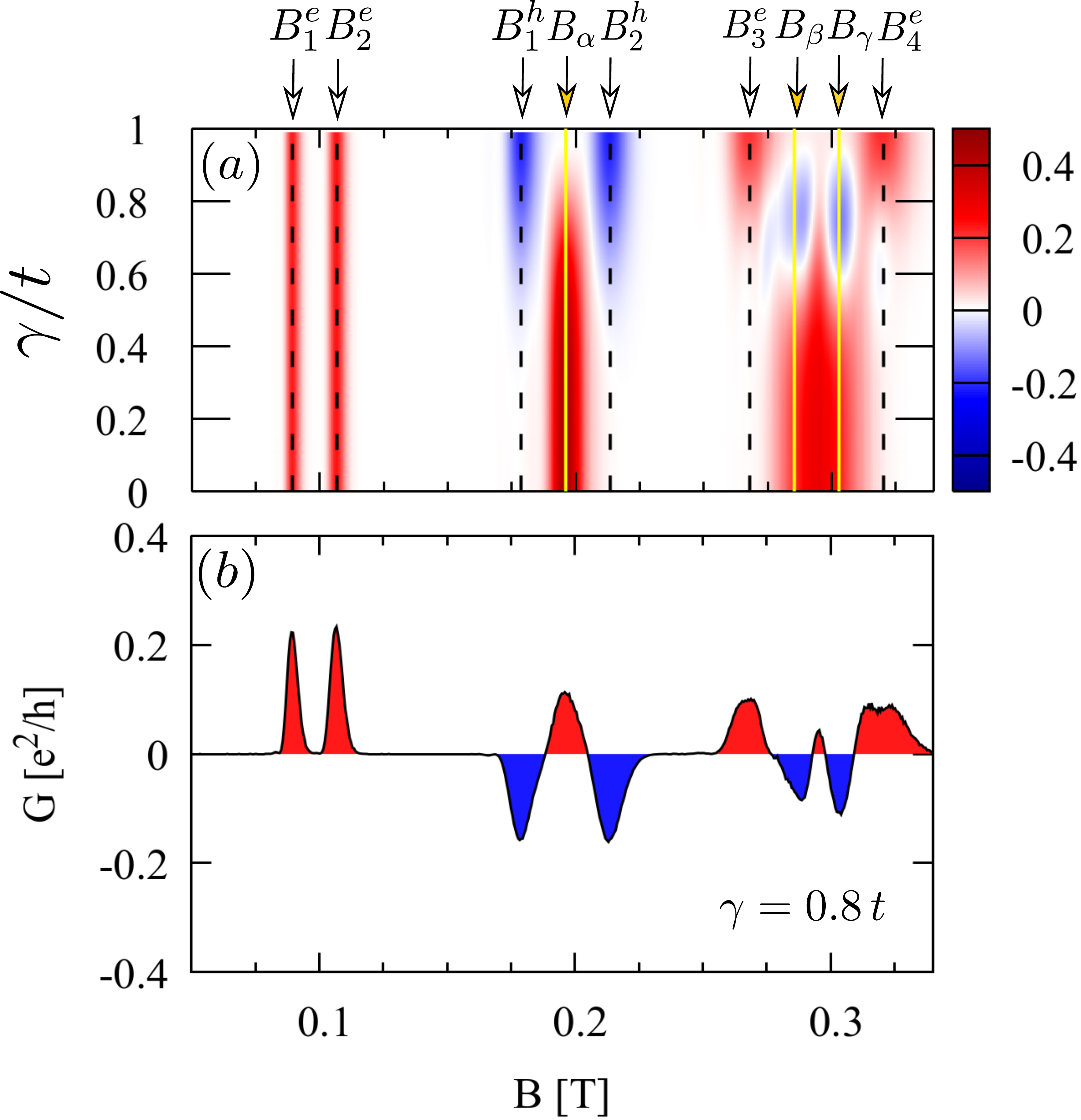}
\caption{(a) Color map of the conductance $G$ between the source and the drain lead as a function of the magnetic field $B$ and the hopping amplitude between the 2DEG and the superconducting terminal $\gamma$ measured in units of $t=\frac{\hbar^2}{2 m^{*}a_0^2}$. (b) Horizontal cut of panel (a) for $\gamma /t=0.8$.}
\label{fig6}
\end{figure}
To better illustrate our findings, we remove the drain contact D and calculate the conductance to an arbitrary point $\bm{r}$ in the sample (where an auxiliary grounded weakly coupled lead is added). In this way, it is possible to build-up a complete image of the ballistic electron-hole cyclotron orbits. We have chosen to work with a large spin-orbit coupling parameter $\alpha=20\,\text{nm eV}$ for the sake of a better visualization of the effect. In Fig.~\ref{fig5} we show these conductance maps for two focusing fields: $B_1^{h}=0.179\,$T [panels I(a)-(c)] and $B_{3}^{e}=0.268\,$T [panels II(a)-(c)]. In panels I(a) and II(a) we only show the electron-electron transmission ($\sum_{\sigma\sigma'}T^{\bm{r}I}_{e\sigma,e\sigma'}$) and, in I(b) and II(b), the electron-hole transmission with its corresponding sign ($-\sum_{\sigma\sigma'}T^{\bm{r}I}_{h\sigma,e\sigma'}$). The total conductance is presented in I(c) and II(c). The region where the superconducting lead is attached is indicated with a black line on the $x$ axis while the injector lead I is delimited with a brown one. The drain lead D, although not taken into account for these calculations, was centered at $249\,a_0$ for the simulations of Fig.~\ref{fig2}. The two cyclotron radii $r_{\pm}$ can be clearly observed for both magnetic fields. The expected semiclassical trajectories associated to each orbit are indicated with dashed (small radius) and solid (large radius) lines. We note that the spatial splitting between both orbits is field dependent: the initial path difference between both trajectories projected along the $x$-direction may be obtained in a semiclassical picture as $\Delta x(B) = 2(r_{+}-r_{-})= \frac{4\alpha m^{*}c}{\hbar e B}$. This formula leads to $\Delta x(B_{1}^{h})\simeq 24 a_0$ and $\Delta x(B_{3}^{e})\simeq 16 a_0$, in good agreement with the splitting observed in the numerical simulations. The linear scaling of the spatial splitting between the focal points at the edge of the sample is also evident from the figure. Due to the good transparency between the sample and the superconductor, the scattering along this interface is dominated by the Andreev reflection channel: incident electrons are entirely scattered as holes and vice versa. This can be seen in the almost perfect cancellation of the electron-electron and electron-hole transmission coefficients near the boundary of the anomalous terminal in panels I(c) and II(c). Being this the case, the two types of semiclassical Andreev edge states keep their orbital radius until a normal scattering event takes place outside the superconducting region.

\begin{figure*}[t]
\includegraphics[width=0.9\textwidth]{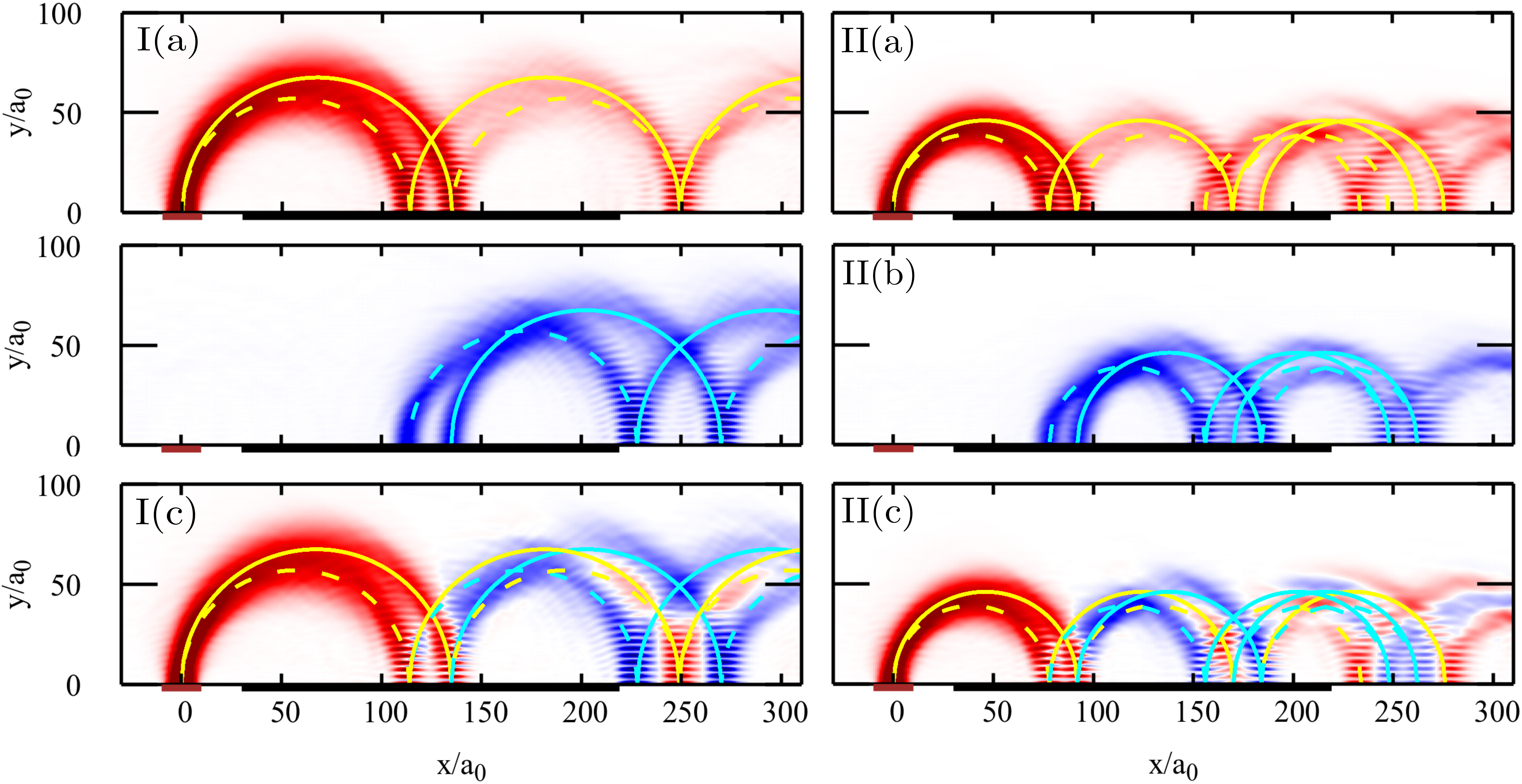}
\caption{Same as Fig.~\ref{fig5} but with a hopping amplitude between the sample and the superconductor of $\gamma=0.8\,t$. In panels I(a)-(c) the magnetic field is chosen to be $B_{\alpha}=0.196\,$T and in panels II(a)-(c) $B_{\beta}=0.285\,$T.}
\label{fig7}
\end{figure*}

A less idealistic scenario can be simulated by lowering the transparency between the superconductor and the 2DEG. This effectively increases the normal scattering probability, and thus provides the possibility to assess the interplay between ordinary and Andreev scattering events. In Fig~\ref{fig6}(a) we show the conductance between the source and the drain lead for different values of the hopping amplitude $\gamma$ between the sample and the superconducting contact. We have kept the parameters of Fig.~\ref{fig2}(c), so that for $\gamma/t =1$ we recover this focusing spectrum. For $\gamma=0$ the superconductor is completely detached from the 2DEG, making the edge to behave as an infinite hard wall potential. Dashed black lines indicate the position of the first six focusing peaks observed in the good transparency limit: $B_1^{e}, B_2^{e}, B_1^{h}, B_2^{h}, B_3^{e}$ and $B_4^{e}$. The normal scattering channel produces the appearance of additional peaks, $B_{\alpha}$, $B_{\beta}$ and $B_{\gamma}$ which are marked with solid yellow lines. In Fig.~\ref{fig6}(b) we show the conductance for a hopping amplitude $\gamma=0.8\,t$, where the effects arising from the coexistence of both scattering channels can be clearly observed. The peak located at $B_{\alpha}=(B_{1}^{h}+B_{2}^{h})/2=0.196\,$T corresponds to the situation where the injected electrons have one intermediate bounce at the edge of the sample. In Fig.~\ref{fig7}, we show for this particular field the electron-electron [panel I(a)] and electron-hole [panel I(b)] transmission coefficients to arbitrary points in the sample.  As clearly seen in panel I(a), the electron-electron scattering leads to semiclassical trajectories with alternating radius so that the $B_{\alpha}$ conductance peak is generated by the focusing condition $2(r_{+}+r_{-})=L$. As observed in panel I(b), Andreev scattering is also present for this particular transparency. This leads to a total conductance [panel I(c)] with an electronic focal point which is accompanied by two hole-like side peaks. As the magnetic field increases, the combination of normal and Andreev scattering events leads to a more involved spectrum. The peaks $B_{\beta}$ and $B_{\gamma}$ are the product of trajectories with two bounces at the superconducting terminal. For low transparencies these are purely electronic-like and can be simply understood as the outcome of two normal scattering events. On the other hand, for better transparencies, the Andreev reflection channel gains weight and the conductance for these fields changes sign ($\gamma \gtrsim 0.6 t$). To understand this behavior, multiple scattering possibilities should be analyzed. In panels II(a)-(c) of Fig.~\ref{fig7} we map the possible electron-hole orbits in real space for the magnetic field $B_{\beta}=0.258\,$T and $\gamma=0.8\,t$. The focusing condition that gives rise to this conductance peak is given by $4r_{-}+2{r}_{+}=L$. The electron-hole transmission coefficient shown in panel II(b) makes clear that the hole-like dip for this particular field has contributions from two different trajectories. In one of them, an electron travelling through the $\mathcal{O}_{-}$ path has a first scattering event dominated by the anomalous channel, so that the Andreev scattered hole preserves the $r_{-}$ radius until it reaches the edge of the sample. If a second ordinary scattering event takes place, the normal-reflected hole will then follow the $\mathcal{O}_{+}$ path. The other contributing trajectory starts from an electronic state travelling through the $\mathcal{O}_{+}$ orbit that has a normal scattering event at first and thus changes its motion to the $\mathcal{O}_{-}$ orbit. In the second scattering event, the electronic state is Andreev-reflected to a hole that keeps the incoming $r_{-}$ radius until it focuses in the drain lead. The hole-like dip in the conductance is then the product of trajectories with two scattering events of different nature: one through the normal channel and one through the Andreev channel. A similar combination explains the $B_{\gamma}$ dip. For $\gamma=t$ these peaks essentially disappear due to the vanishing of the normal scattering probability.

\section{Summary}\label{IV}

We have numerically studied the transport properties of a transverse magnetic focusing device as the one illustrated in Fig.~\ref{fig1}, where two lateral normal contacts are attached to a 2DEG with a third superconducting terminal in between. In the weak or moderate field regimes, with large filling fractions, injected carriers travel along bouncing cyclotron orbits alternating their electron and hole nature due to Andreev scattering at the superconducting lead. We have analyzed how the presence of Rashba spin-orbit coupling affects the cyclotron motion of these semiclassical chiral Andreev edge states by studying spin and charge transport in the device. The spin-orbit interaction leads to the unfolding of the Fermi surface into two pieces with different spin texture which are characterized by the wavevectors $k_{\pm}$. The semiclassical skipping trajectories along the boundary of the sample are accordingly spin-split into two nonidentical paths, differentiated by their orbital radius $r_{\pm}$.  Our work, including numerical simulations an analytical estimations, emulates this situation.

Spin separation in cyclotron motion is a well-known phenomenon which has been thoroughly studied in the absence of a superconducting edge. In such cases, the normal spin-conserving scattering at the sample edge imposes a transition of the scattering states from one Fermi surface piece to the other and consequently mixes the two orbital paths. The peculiarity of the hybrid Hall/superconductor interface is the opening of the Andreev reflection channel. This type of scattering goes in hand with the transport of singlet Cooper pairs between the sample and the anomalous terminal: an incoming electron (hole) with a given spin is reflected as a hole (electron) with the opposite spin. In this way, Andreev scattering connects antipodal points in the same piece of the Fermi surface, preserving the band index. As the two bands ($\pm$) give rise to semiclassical orbits with a different but well defined radius ($r_{\pm}$), the two spin dependent trajectories in real space separate more the greater the number of bounces at the hybrid interface. This effect can be measured in the conductance spectra as a splitting of the focusing peaks that increases with the peak index up to a point where different peaks overlap and mix. The amplification of the spin-split focal points could be a handy tool to improve the ability to selectively separate spin-polarized currents in these devices. At the same time, the alternating electron-hole nature of the chiral states allows for the possibility to control the carriers charge. 
We supported our results by calculating normal and anomalous transmissions to arbitrary points in the sample. These transport measurements provide a means of building a complete map of the electron-hole cyclotron orbits, which nicely compare with our semiclassical picture.

To describe situations more in line with experimental conditions, we have analyzed the case of non-ideal couplings between the sample and the superconducting terminal. This was done by lowering the transparency of the junction. For large transparencies ($\gamma / t \approx  0.8$) the first focusing peaks observed in the ideal case are recovered together with additional peaks that stem from the normal scattering channel. For low transparencies ($\gamma/t <0.4$) the anomalous Andreev scattering is significantly reduced and the bouncing chiral carrier recovers its pure electron-like nature: all the conductance peaks become positive. The focusing fields and the intensity of the focusing peaks can then provide direct information on both the magnitude of the SO coupling and the quality of the junction between the sample and the superconductor.

The recent development of high-quality hybrid devices based on two dimensional electron gases in contact with superconductors will pave the road for new experiments to come, where a considerable interplay between strong SO coupling and superconducting correlations should be taken into account (as in the case of In or Sb based semiconducting heterostructures~\citep{Zhi2019}). Our results, although simple in nature, show how the coherent ballistic paths of chiral Andreev edge states can be used to engineer and control charge and spin transport in these novel platforms. 

\begin{acknowledgments}
We acknowledge financial support from ANPCyT (grants PICT 2016-0791 and PICT 2018-1509), CONICET (grant PIP 11220150100506) and SeCyT-UNCuyo (grant 06/C603).
\end{acknowledgments}

\end{document}